\documentclass[a4paper,preprintnumbers,floatfix,
superscriptaddress,prl,twocolumn,showpacs,
notitlepage,longbibliography]{revtex4-1}

\usepackage[utf8]{inputenc}
\usepackage{graphicx}
\usepackage{dcolumn}
\usepackage{bm}
\usepackage{braket}

\begin{document}


\title{Experimental realization of a quantum CNOT gate for orbital angular momentum and polarization with linear optical elements}



\author{J. H. Lopes}
\affiliation{%
 Grupo de F\'isica da Mat\'eria Condensada, N\'ucleo de Ci\^encias Exatas - NCEx, Campus Arapiraca, Universidade Federal de Alagoas, 57309-005 Arapiraca-AL, Brazil
}%

\author{W. C. Soares}
\affiliation{%
 Grupo de F\'isica da Mat\'eria Condensada, N\'ucleo de Ci\^encias Exatas - NCEx, Campus Arapiraca, Universidade Federal de Alagoas, 57309-005 Arapiraca-AL, Brazil
}%
\affiliation{%
Departamento de Física, Universidade Federal de Santa Catarina, Florianópolis, SC, 88040-900, Brazil
}%
\author{Bert\'ulio de Lima Bernardo}
 \affiliation{Departamento de F\'isica, Universidade Federal da Para\'iba, Caixa Postal 5008, 58051-900 Jo\~ao Pessoa, PB, Brazil}
 
\author{D. P. Caetano} 
\affiliation{%
Universidade Federal Fluminense, Escola de Engenharia Industrial Metalúrgica de Volta Redonda.
Avenida dos Trabalhadores, 420
Vila Santa Cecília
27255125 - Volta Redonda, RJ - Brasil
}%

\author{Askery Canabarro}
\email{askery.canabarro@arapiraca.ufal.br}
\affiliation{%
 Grupo de F\'isica da Mat\'eria Condensada, N\'ucleo de Ci\^encias Exatas - NCEx, Campus Arapiraca, Universidade Federal de Alagoas, 57309-005 Arapiraca-AL, Brazil
}%
\affiliation{%
 International Institute of Physics, Federal University of Rio Grande do Norte, 59070-405 Natal, Brazil
}%




\date{\today}

\begin{abstract}
We propose and experimentally demonstrate that a Mach-Zehnder interferometer composed of polarized beam splitters and a pentaprism in the place of one of the mirrors works as a linear optical quantum controlled-NOT (CNOT) gate. To perform the information processing, the polarization and orbital angular momentum (OAM) of the photons act as the control and target qubits, respectively. The readout process is simple, requiring only a linear polarizer and a triangular diffractive aperture before detection. The viability and stability of the experiment makes the present proposal a valuable candidate for future implementations in optical quantum computation protocols.     
\end{abstract}


\maketitle


\section{\label{sec:1}Introduction}

The capacity of a photon to carry information can be greatly increased if one accesses its many degrees of freedom: spatial, polarization, frequency and the orbital angular momentum (OAM). In doing so, it is possible to make a single-photon encode multiple qubits, whose manipulation can give rise to the implementation of photonic quantum logic gates, which are the building blocks of future technologies such as quantum communication \cite{nielsen,ekert,bennett,bennett2} and  quantum computation \cite{knill,bow,shor,lany}. The great advantage in using light for such purposes is the fact that photons are practically unaffected by detrimental decohering processes \cite{zurek,schloss,bert,dilsonOL,dilsonBJP,dilsonJPB}. A celebrated result in quantum information science was the demonstration that single-qubit and controlled-NOT (CNOT) gates together are sufficient for realizing universal quantum computation \cite{vince}. In realizing such task with optical schemes, the information is commonly encoded on the polarization degree of freedom, in which the one-qubit logic gates are carried out with birefringent waveplates \cite{kok}. However, the implementation of an optical two-qubit CNOT operation for a single-photon demands necessarily the control of a second degree of freedom \cite{kok2}.   

In 2001, Knill, Laflamme and Milburn \cite{knill} demonstrated a probabilistic method to realize efficient quantum computation, which is efficient and scalable, using only linear optical elements, photodetectors and single-photon sources. After that, a number of works have proposed ways to construct CNOT gates with linear optics. For example, Fiorentino et al. created a linear CNOT logic gate using the polarization and momentum of a single photon \cite{fio}. Oliveira et. al. realized a single-photon CNOT gate manipulating the polarization and the transverse spatial modes of the electromagnetic field \cite{oli}. Deng et. al. theoretically proposed a CNOT operation involving the polarization and the OAM of a single-photon, but the use of computer generated hologram (CGH) inside the circuit was required \cite{deng}. More recently, an experimental realization of a photonic CNOT gate also involving polarization and OAM was reported, however, making use of nonlinear optical elements \cite{zeng}.   

In this work, we theoretically study and experimentally verify an optical implementation of a CNOT logic gate manipulating the polarization and OAM degrees of freedom only with linear optical elements. The polarization orientation and the sign of the OAM state are the control and target qubits, respectively. Contrary to the theoretical proposal by Deng and co-workers \cite{deng}, our CNOT scheme does not require the usage of CGH to process information, which undoubtedly is a practical advantage towards a scalable construction of quantum information processors. We also show that the same apparatus can generate the family of maximally entangled Bell states involving polarization and OAM states. The readout process, which usually is not straightforward for OAM states, could be performed through the pragmatic triangular aperture method \cite{jandir}, therefore, preventing faulty measurement outcomes.

\section{\label{sec:2}Theory}

It is well known that light beam with the phase dependence $\exp(i \ell \phi)$, where $\phi$ is the azimuthal variable, carries an OAM of $\ell \hbar$ per photon, with $\ell$ integer \cite{allen,yao}. This property, which takes place when the beam wavefront has a helical structure, represents a new photonic degree of freedom that raised the possibility of using its high-dimensionality to encode a large amount of information in a single photon \cite{barreiro,malik,fick}. Particularly, the joint manipulation of polarization and OAM to access a higher-dimensional quantum state space of a unique photon is progressively allowing the realization of novel optical quantum information protocols \cite{yao}. In the same line of thought, in this section we propose an experimental method to use OAM states along with polarization to construct a linear optical CNOT gate.

Our experimental proposal consists in submitting a light beam endowed with an arbitrary nonzero OAM to a modified Mach-Zehnder interferometer (MZI) composed by two polarizing beam splitter (PBS), in which horizontally- and vertically-polarized fields are respectively transmitted and reflected, a mirror, a pentaprism, and a charge-coupled-device (CCD) camera. In the experimental setup, which is sketched in Fig.1, the light beam will have the horizontally and vertically-polarized components separated by the first polarizing beam splitter, PBS$_{1}$. The former will be deflected by 90$^\circ$ towards the second beam splitter PBS$_{2}$, while the latter will reflect twice inside the pentaprism before being deviated. The net effect of the pentaprism is to deviate the light beam by 90$^\circ$ with high precision, facilitating the alignment of the interferometer, and invert the image, due to the double reflection, as described in Ref. \cite{sasada}. This last fact causes an inversion in the OAM sign of the output beam that is also sent towards PBS$_{2}$.

\begin{figure}[ht]
\centerline{\includegraphics[width=7.5cm]{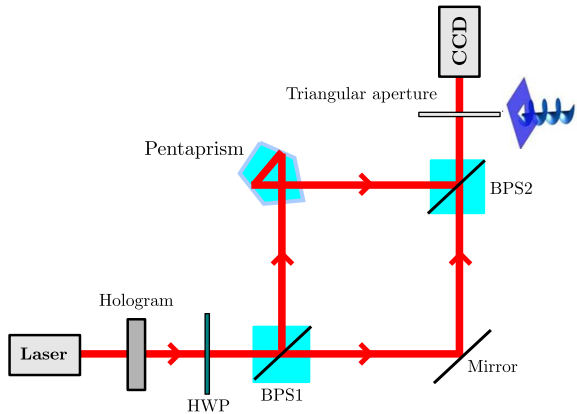}}
\caption{Sketch of the optical circuit used to implement a quantum CNOT gate with the polarization and the OAM states acting respectively as the control and target qubits.}
\label{setup}
\end{figure}

In order to understand the effect of the proposed optical circuit, let us investigate, for example, the transformations obtained in the evolution of an input beam that is horizontally-polarized and has an arbitrary OAM quantified by $\ell$. After being transmitted through PBS$_{1}$, the beam will be deflected in the mirror and sent towards PBS$_{2}$ to be again transmitted and detected by the CCD camera. Note that the states of both polarization and OAM are unchanged in this lower arm of the interfe\-rometer. On the other hand, in considering a vertically-polarized input beam with an arbitrary OAM, $\ell$, it will be reflected at PBS$_{1}$ to be later deviated by the pentaprism with an inverted OAM, $-\ell$. This beam will then suffer a new reflection at PBS$_{2}$ before being detected by the CCD camera. 

Now, let us translate the overall effect of this optical circuit in terms of quantum logic operations. First, we adopt some basis states for the representation of polarization and OAM. The horizontally- and vertically-polarized states are $\ket{0}_{p}$ and $\ket{1}_{p}$, respectively. In turn, positive and negative OAM states will be represented respectively by $\ket{0}_{0}$ and $\ket{1}_{0}$. In this form, the basis states that span our four-dimensional space with the polarization and OAM of the photons are $\{\ket{0_{p}0_{0}},\ket{0_{p}1_{0}},\ket{1_{p}0_{0}},\ket{1_{p}1_{0}} \}$. Observe that, if we consider the polarization states as the control and the OAM state as the target, with the experimental proposal above, the four classical CNOT gate transformations are naturally implemented,
\begin{equation}
\label{1}
\ket{0_{p}0_{0}} \rightarrow \ket{0_{p}0_{0}},
\end{equation}
\begin{equation}
\label{2}
\ket{0_{p}1_{0}} \rightarrow \ket{0_{p}1_{0}},
\end{equation}
\begin{equation}
\label{3}
\ket{1_{p}0_{0}} \rightarrow \ket{1_{p}1_{0}},
\end{equation}
\begin{equation}
\label{4}
\ket{1_{p}1_{0}} \rightarrow \ket{1_{p}0_{0}}.
\end{equation}
However, the optical circuit of Fig. 1 also serves as a quantum CNOT logic gate. Indeed, if one considers a general combination of the basis states, 
\begin{eqnarray}
\label{5}
\ket{\psi} = a\ket{0_{p}0_{0}}+b\ket{0_{p}1_{0}}+c\ket{1_{p}0_{0}}+d\ket{1_{p}1_{0}}, 
\end{eqnarray}
the following transformation takes place:
\begin{eqnarray}
\label{6}
\ket{\psi^{'}} \rightarrow a\ket{0_{p}0_{0}}+b\ket{0_{p}1_{0}}+c\ket{1_{p}1_{0}}+d\ket{1_{p}0_{0}},
\end{eqnarray}
where the coefficients $a$, $b$, $c$ and $d$ are arbitrary complex numbers, satisfying the normalization condition, $|a|^{2}+|b|^{2}+|c|^{2}+|d|^{2}=1$. Eq.~\ref{6} is the necessary condition for the implementation of a quantum CNOT gate, i.e., if the control qubit is 0 (1), the target qubit is unchanged (changed).

Another interesting aspect related to the circuit of Fig. 1 is the possibility of creating entangled states involving the polarization and the OAM degree of freedom of a single-photon. This task can be realized if a Hadamard gate is applied to the control qubit (polarization) prior to the quantum CNOT gate \cite{nielsen}. In fact, with this procedure we are able to create maximally entangled Bell states involving the polarization and OAM Hilbert spaces. Therefore, what we need here is an optical element that acts as a Hadamard operation in the polarization state space. A waveplate placed before PBS$_{1}$ that rotates the polarization by 45$^{\circ}$, would play this role. In this case, in considering both the polarization and OAM states, if the polarization of the incident light is horizontal (vertical), after the waveplate, it is transformed into a diagonal (antidiagonal) polarization, that is,   
\begin{equation}
\label{7}
\ket{0_{p}0_{0}} \rightarrow \frac{1}{\sqrt{2}}(\ket{0_{p}0_{0}}+\ket{1_{p}0_{0}}),
\end{equation}
\begin{equation}
\label{8}
\ket{0_{p}1_{0}} \rightarrow \frac{1}{\sqrt{2}}(\ket{0_{p}1_{0}}+\ket{1_{p}1_{0}}),
\end{equation}
\begin{equation}
\label{9}
\ket{1_{p}0_{0}} \rightarrow \frac{1}{\sqrt{2}}(\ket{0_{p}0_{0}}-\ket{1_{p}0_{0}}),
\end{equation}
\begin{equation}
\label{10}
\ket{1_{p}1_{0}} \rightarrow \frac{1}{\sqrt{2}}(\ket{0_{p}1_{0}}-\ket{1_{p}1_{0}}).
\end{equation}
Therefore, after these operations, the transformed state will be submitted to the circuit of Fig. 1. The circuit will execute the CNOT operation given by the rules of Eqs.~\ref{1} to~\ref{4}, producing the complete family of maximally entangled Bell states  
\begin{equation}
\label{11}
\ket{0_{p}0_{0}} \rightarrow \frac{1}{\sqrt{2}}(\ket{0_{p}0_{0}}+\ket{1_{p}1_{0}}),
\end{equation}
\begin{equation}
\label{12}
\ket{0_{p}1_{0}} \rightarrow \frac{1}{\sqrt{2}}(\ket{0_{p}1_{0}}+\ket{1_{p}0_{0}}),
\end{equation}
\begin{equation}
\label{13}
\ket{1_{p}0_{0}} \rightarrow \frac{1}{\sqrt{2}}(\ket{0_{p}0_{0}}-\ket{1_{p}1_{0}}),
\end{equation}
\begin{equation}
\label{14}
\ket{1_{p}1_{0}} \rightarrow \frac{1}{\sqrt{2}}(\ket{0_{p}1_{0}}-\ket{1_{p}0_{0}}),
\end{equation}
which demonstrate the entangling properties of the proposed experimental setup. 

A fundamental stage in implementing any quantum information protocol is obviously the readout process. In the present case, it is essential that the four basis states of Eqs.~\ref{1} to~\ref{4} be distinguishable in an experimental rea\-lization. The readout of the polarization basis states, $\ket{0}_{p}$ and $\ket{1}_{p}$, is trivial with the usage of linear polari\-zers oriented along the horizontal and vertical directions. Nevertheless, the readout of the OAM basis states, $\ket{0}_{0}$ and $\ket{1}_{0}$, is not so straightforward. Indeed, by direct detection of a light beam endowed with OAM, one can only see the well-known donut-shaped intensity profile, independent of the topological charge of the optical vortex \cite{yao}. 

To circumvent this problem, we propose the observation of the far field pattern of the output beam after being diffracted by a triangular aperture. This method was proposed in Ref. \cite{jandir} as a practical and precise procedure to determine the OAM of optical fields. Specifically, it was found that the diffraction pattern is a triangular optical lattice whose number of bright points $N$ on any side of the triangle is related with the topological charge as $\ell=N-1$. However, the readout process in our protocol does not require the knowledge about $\ell$, but only its sign. This is where the triangular aperture method is more effective. The sign of $\ell$, which is related to the handedness of the optical vortex, is found out just by observing the direction to which the triangular optical lattice is pointing. If the triangular lattice points to same direction of the triangular aperture is because the the sign of the OAM is positive, which means $\ket{0}_{0}$. Conversely, if the triangular lattice points to the opposite direction, it signifies that the OAM sign is negative, rendering $\ket{1}_{0}$. In this form, for the present setup, the triangular aperture should be placed before the CCD camera for the observation of the handedness of the output optical vortex.                      

\section{Experimental methods and results}

\begin{figure}[htb]
\begin{center}
\includegraphics[height=2.5in]{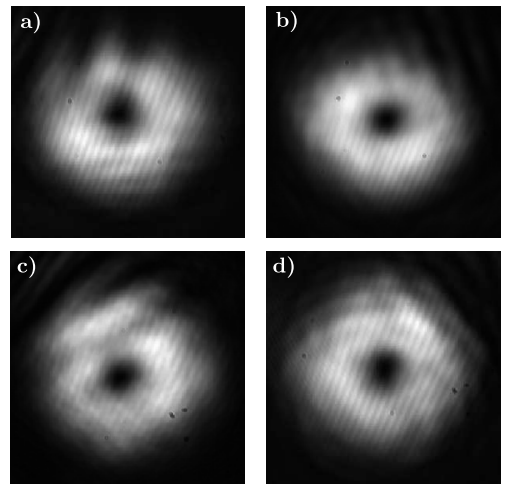}
\end{center}
\label{F2}
\caption{Experimental intensity profile of the four output states of the setup of Fig. 1, a) $\ket{0_{p}0_{0}}$, b) $\ket{0_{p}1_{0}}$, c) $\ket{1_{p}0_{0}}$, d) $\ket{1_{p}1_{0}} $, by using a linear polarizer and the triangular aperture prior to detection to realize the readout process. The four states cannot be distinguished.}
\end{figure}

\begin{figure}[htb]
\begin{center}
\includegraphics[height=2.5in]{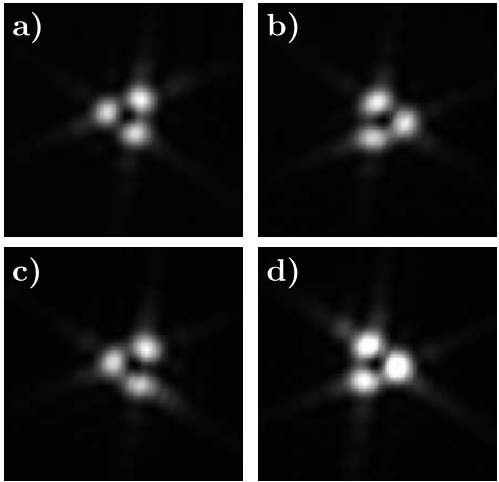}
\end{center}
\label{F3}
\caption{Experimental intensity profile of the four output states of the setup of Fig. 1, a) $\ket{0_{p}0_{0}}$, b) $\ket{0_{p}1_{0}}$, c) $\ket{1_{p}0_{0}}$, d) $\ket{1_{p}1_{0}} $, by using a linear polarizer and the triangular aperture prior to detection to realize the readout process. The four states could be naturally distinguished.}
\end{figure}

We realized an experiment following the theoretical proposal sketched in Fig. 1. The light source used was the Gaussian mode of an Argon Laser operating at 532 nm with a power of 10 mW with vertical polarization. The beam then illuminates a CGH with controllable pixels written in an (Hamamatsu Model X10468-01) spatial light modulator (SLM) for the generation of the helical phase profile, $\exp (i \ell \phi)$. In our experiment, we created a beam with $\ell = 1$ before PBS$_{1}$. Next, the reflected beam is deviated by the pentaprism, which also inverts the sign of the OAM, to recombine with the beam transmitted by PBS$_{1}$ at the second polarized beam splitter PBS$_{2}$. After this stage, the output beam passes through an equilateral triangular aperture with side length of 2 mm. Then, a 30 cm focal length lens was placed immediately after the aperture to create the far field diffraction pattern at the focal length of the lens. Finally, the intensity diffraction pattern is registered by the CCD camera. 

The preparation and measurement of the control qubit (polarization) is made by using a half-wave plate (HWP) before PBS$_{1}$, if one wants to make the input polarization horizontal, and a linear polarizer placed after PBS$_{2}$ oriented either along the vertical or the horizontal direction in order to verify the final control state. 
\begin{table}
  \centering
  \caption{Truth Table}\begin{tabular}{cc} \\ \hline
    Input state & Output state \\ \hline
    H polarization; $l=1$  & H polarization; $l=1$  \\
    H polarization; $l=-1$ & H polarization; $l=-1$  \\
   V polarization; $l=-1$ & V polarization; $l=1$  \\
   V polarization; $l=1$ & V polarization; $l=-1$  \\ \hline
  \end{tabular}
\end{table}
The target (OAM) state is measured according to the orientation of the triangular truncated optical lattice observed in the camera \cite{jandir}. In order to see the efficiency of the method, in Fig. 2 we show the intensity patterns of the four output states of Eqs.~\ref{1} to \ref{4} recorded by the camera when the triangular aperture is absent in the experiment. As can be seen, all we obtain is the characteristic donut-like intensity profile, without OAM state information. In a different perspective, Fig. 3 shows the pattern when the aperture is present. We observe that the states with $\ell = 1$ and $\ell = -1$ are easily distinguishable. Table I shows the truth table of our CNOT gate, whose output state profiles are shown in Fig. 3. The results completely agree with the transformations indicated in Eqs.~\ref{1} to~\ref{4}.    

\section{Conclusion}

To sum up, we theoretically and experimentally demonstrated a linear optical quantum CNOT gate involving the polarization and OAM degrees of freedom, which serve as the control and target qubits, respectively. The experimental setup does not demand the usage of computer generated holograms to process information; a fact that makes the arrangement suitable for scaling up quantum-computing architectures. The readout process is based only on the use of a linear polarizer and a diffractive triangular aperture. With the present configuration, the experimental results confirmed our theoretical predictions in an unambiguous way. Given the stability of the experimental setup and the robustness of the polarization and OAM states against environmental noise, we consider the present proposal a promising candidate to be used in future schemes of linear optical quantum computation.



\section*{Acknowledgements}

The authors acknowledge the Brazilian funding agencies CNPq (AC's Universal grant No. 423713/2016-7 \& BLB's PQ grant No. 309292/2016-6), CAPES  and Alagoas state research agency FAPEAL. We also thank  
UFAL (AC's and WCS's paid license for scientific cooperation at UFRN and UFSC, respectively) and
MEC/UFRN (AC's postdoctoral fellowship).

\bibliography{mybibfile}


\end{document}